\documentclass[12pt]{iopart}

\usepackage{hyperref}
\usepackage{graphicx}
\usepackage{srcltx}

\begin{document}

\title[Future Experimental Programs Murayama]{Future Experimental Programs}

\author{Hitoshi Murayama}
\address{Department of Physics, University of California,
  Berkeley, California 94720, USA\\
  Theoretical Physics Group, Lawrence Berkeley National
  Laboratory, Berkeley, California 94720, USA\\
  Kavli Institute for the Physics and Mathematics of the
  Universe (WPI), Todai Institutes for Advanced Study, University of Tokyo,
  Kashiwa 277-8583, Japan} 
\ead{hitoshi@berkeley.edu, murayama@lbl.gov, hitoshi.murayama@ipmu.jp}
\begin{abstract}
  I was asked to discuss future experimental programs even though I'm
  a theorist.  As a result, I present my own personal views on where
  the field is, and where it is going, based on what I myself have
  been working on.  In particular, I discuss why we need {\it
    expeditions}\/ into high energies to find clues to where the
  relevant energy scale is for dark matter, baryon asymmetry, and
  neutrino mass.  I also argue that the next energy frontier machine
  should be justified on the basis of what we know, namely the mass of
  the Higgs boson, so that we will learn what energy we should aim at
  once we {\it nail}\ the Higgs sector.  Finally I make remarks on
  dark energy.
\end{abstract}

\maketitle

\section{Introduction}

The discovery of a ``Higgs-like particle'' on July 4, 2012 was a truly
historic moment in the history of science
\cite{Aad:2012tfa,Chatrchyan:2012ufa}.  Many of us in the United
States watched the seminar at CERN over webcast in the midnight hours.
Given that it was announced on the Independence Day of the United
States, we celebrated the {\it Higgsdependence}\/ Day in the early
morning.

So far, what we've seen looks {\it minimal}\/.  Later, a CERN
announcement made in March 2013 said it is {\it a}\/ Higgs boson.
Indeed, the newly discovered particle looks very much like {\it the}\/
Standard Model Higgs boson.  We've been after this particle ever since
1933 when Fermi wrote his theory of nuclear beta decay.  There, he
introduced a constant $G \approx 10^{-5} m_p^{-2}$ which we now call
the Fermi constant $G_F$.  It corresponds to the energy scale
$G_F^{-1/2} \approx 300$~GeV, and we learned from him that something
is going on at this energy scale.  It took a whopping eighty years to
come to the point where we now have a UV-complete theory of strong,
weak, and electromagnetic forces with all of the parameters measured.
In fact, it is a renormalizable and consistent theory that may be
valid all the way up to the Planck scale.  Coincidentally, even
cosmology looks minimal given the Planck data \cite{Ade:2013uln},
which suggests a minimal single-field inflation.  Maybe the year 2013
will be remembered in history as {\it the year of elementary
  scalars}\/.

\begin{figure}
\begin{center}
\includegraphics[clip,width=.35\textwidth]{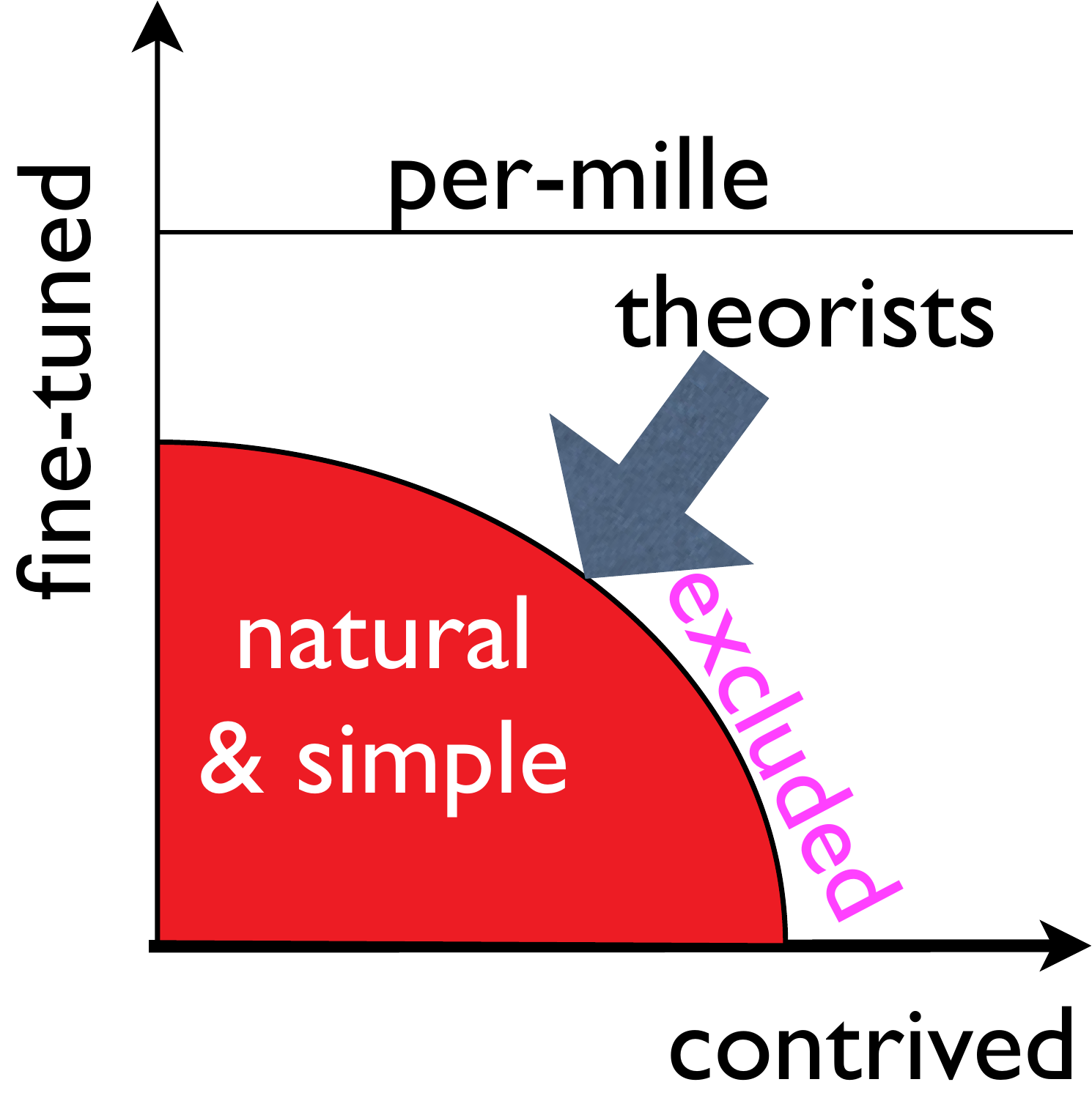}
\end{center}
\caption{Schematic constraints on space of theories.}
\label{fig:finetuning}
\end{figure}

Despite this achievement, or rather because of it, there is a building
anxiety in the community.  How come we don't see anything else?  Will
the progress stop?  There is no sign of physics beyond the Standard
Model in the LHC data.  For a typical search for supersymmetric
particles, for example, squarks and gluinos are excluded up to 1.3~TeV
or so.  On the other hand, the conventional arguments based on the
naturalness concept suggested that we should have new particles that
stabilize the electroweak scale below TeV.  It appears that ``natural
and simple'' models have been excluded (Fig.~\ref{fig:finetuning}).
Then we have two directions to go in: the less natural, namely
fine-tuned, or the less simple, namely contrived.  At the same time,
theorists are trying to come up with models that can evade the current
experimental limits, pushing back on this problem.  See
\ref{sec:appendix} for my own recent attempts.

I have to point out, however, that certain levels of fine-tuning do
occur in nature.  All examples I'm aware of, with the glaring
exception of the cosmological constant, are at most at the level of a
few per-mille.  The current LHC limit has not quite reached that level;
the next runs at 13--14~TeV may well reveal new physics as we hoped
for.  I will come back to this question later in this talk.

In any case, it is true that experimental limits have started to haunt
theorists.  Theorists used to complain that the experiments had a hard
time keeping up with their new ideas.  Now the tide has reversed.
Theorists are being {\it threatened}\/ by new data.  I believe this is
quite a healthy field!

Nonetheless, having a fully UV-complete theory of the Minimal Standard
Model, now supported by the new particle that has been discovered,
makes us ask the following question:

\section{Is particle physics over?}

On this question, fortunately, the answer is a resounding {\it no}\/.
Since 1998, we have discovered five pieces of empirical evidence for
physics beyond the Standard Model thanks to tremendous progress in
experiments.

First, non-baryonic dark matter.  Even though dark matter had been
discussed since 1930's by Fritz Zwicky, it was not clear whether dark
matter would be dark astronomical objects or hidden baryons.  This
issue was completely settled in 2003.  The search for dark
astronomical objects (MACHOs = Massive Compact Halo Objects) excludes
the possibility that Galactic halo consists solely of MACHOs between
about $\times 10^{-7} M_\odot$ and $10 M_\odot$ \cite{Afonso:2002xq}.  On
the other hand, the power spectrum in the Cosmic Microwave Background
(CMB) anisotropy by WMAP (Wilkinson Microwave Anisotropy Probe)
excludes the baryonic dark matter completely as a discrepancy between
the overall matter density $\Omega_M h^2 = 0.14 \pm 0.02$ and the
baryon density $\Omega_b h^2 = 0.024 \pm 0.001$ \cite{Spergel:2003cb}.
We are learning what dark matter is not, but not what it is.  In fact,
we know so little that the only model-independent lower limit on the
dark matter mass comes from the requirement that its ``Bohr radius''
in the gravitational potential must fit within the galactic scale
\cite{Hu:2000ke}.  Combined with the MACHO search, we managed to
narrow down its mass from $10^{-31}$~GeV to $10^{50}$~GeV, {\it
  i.e.}\/, to within 81 orders of magnitude.  Zwicky must be happy to
see our progress!  Thus, we need to keep our minds {\it very}\/ open
about the nature of dark matter.

The flavor oscillation of neutrinos, and hence their finite masses, is
not a part of the Minimal Standard Model either, arguably the first
established physics beyond the Standard Model in 1998
\cite{Fukuda:1998mi}, revealing the mixing angle $\theta_{23}$.  Later
on, the oscillation (or rather lack of it as a result of the matter
effect) of solar neutrinos \cite{Ahmad:2002jz} and oscillation of
reactor neutrinos \cite{Eguchi:2002dm} pointed to the same parameter
set (and the angle $\theta_{12}$) in 2002 resolving a puzzle that goes
back half a century.  The final mixing angle $\theta_{13}$ was
discovered in 2012 \cite{An:2012eh}.  Some people think it is
only a minor extension of the Standard Model, but it should be
emphasized that we don't yet know {\it how}\/ it should be extended.

The accelerated expansion of the Universe came as a big surprise to
all of us \cite{Perlmutter:1998np,Riess:1998cb}.  Its cause is now
called dark energy, even though we are very far away from
understanding what it is.  It may be cosmological constant, due to a
miraculous cancellation between quantum fluctuation of the vacuum and
a classical constant energy density for 120 digits.  It may be some
dynamical substance called quintessence.  Either way, it is very
difficult to understand its overall amount.

At the same time, the observed {\it apparently acausal}\/ density
fluctuations in the CMB cannot be explained by the Standard Model.  The
CMB photons that came from one end of the Universe have just reached
us; they seem to be correlated with the CMB photons that came
from the other end, when they have had no chance to meet and set up their
temperatures.  This is what I mean by {\it acausal}\/.  The best
explanation is that they {\it were}\/ in fact in causal contact early
on because the entire visible Universe was much smaller than a
nucleus; it was later stretched to a macroscopic size by an
exponential expansion called inflation.  The latest Planck data
strongly supports this idea \cite{Ade:2013uln}.  We normally assume
that it was caused by a scalar field called the inflaton rolling slowly
down the hill, but we don't know what it is, nor how it couples to the
Standard Model particles.

Finally, once we accept the inflationary paradigm, the cosmic baryon
asymmetry $\eta_b = n_b/n_\gamma \approx 5\times 10^{-10}$ cannot be
assumed to be the initial condition of the Universe.  This is because
the enormous exponential expansion (normally assumed to be more than
$e^{60}$) wipes out any pre-existing asymmetry.  This implies that the
baryon asymmetry needs to be {\it created}\/ after the inflation by a
microphysical process.  On the other hand, the CP violation in the
Standard Model is now known to be incapable of producing enough baryon
asymmetry.  This is because that we now have understood the known CP
violating phenomena by the Kobayashi--Maskawa theory thanks to the
$B$-factory experiments starting in 2001
\cite{Abe:2001xe,Aubert:2001nu}.  This means that the Standard Model
cannot generate the baryon asymmetry larger than the Jarlskog
invariant $J = \Im m (Tr [Y_u^\dagger Y_u, Y_d^\dagger Y_d]) \approx
10^{-20}$ \cite{Jarlskog:1985ht}, further suppressed by small
efficiencies or powers of coupling constants in known mechanisms.

So, it is clear that particle physics is far from over.  There are at
least five important pieces of data that are crying out to be
explained and understood.  The catch is that we don't know the {\it
  energy scale}\/ of physics relevant to these mysteries.  Right now
we are on fishing expeditions. In particular, we are and will continue
to be looking for new phenomena and new sources of CP violation in the
quark sector (LHCb, SuperKEKB, rare kaon decays), lepton sector
(neutrino oscillations, neutrinoless double-beta decay, and electric
dipole moments), and their combination (proton decay).  We try to cast
a wide net, hoping to catch any interesting fish, so that we learn
where the next important energy scale is.  In a sense, this is what
Fermi succeeded in doing; by observing rare phenomena of nuclear
$\beta$-decays, which violate conservation law of neutron and proton
numbers that all other known forces respect, they were caught in the net
and we learned about the Fermi scale.

\begin{figure}
\begin{center}
\includegraphics[clip,width=.85\textwidth]{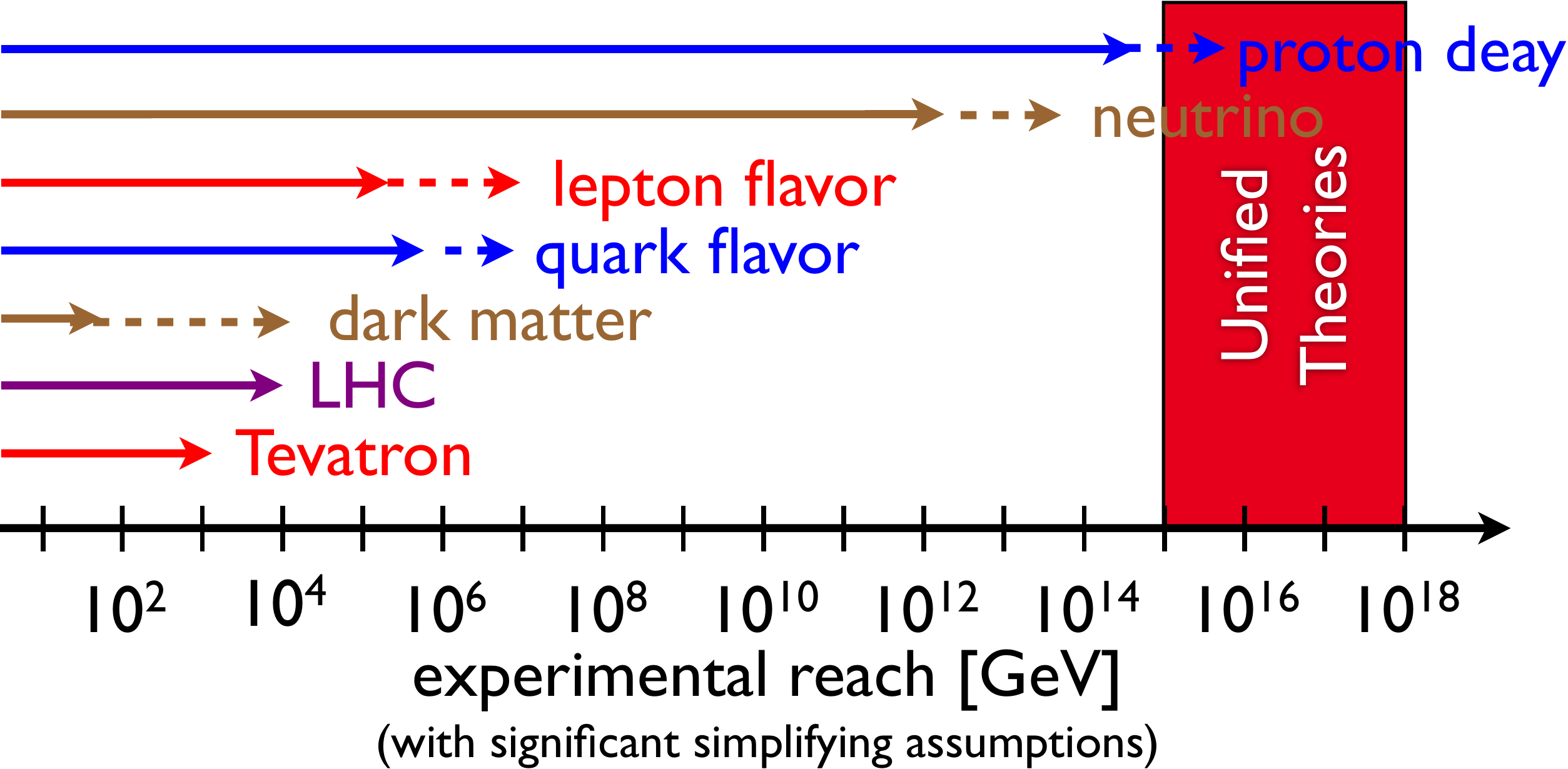}
\end{center}
\caption{Approximate energy reach for expeditions.  Solid arrows
  indicate the current reach, while the dashed arrows anticipated
  improvements by proposed experiments \cite{Ligeti}.}
\label{fig:expedition}
\end{figure}

Whatever the next energy scale beyond the Standard Model is, it plays
the role of the UV cutoff of the Standard Model as a low-energy
effective field theory.  And the effects of the UV cutoff
can be parametrized by higher dimension operators power suppressed by
$\Lambda_{UV}$ added to the Standard Model,
\begin{equation}
  {\cal L} = {\cal L}_{SM} + \frac{1}{\Lambda_{UV}} {\cal L}_5
  + \frac{1}{\Lambda_{UV}^2} {\cal L}_6 + \cdots
\end{equation}
We need to first look at the Standard Model Lagrangian ${\cal L}_{SM}$
which has the structure as shown on T-shirts from CERN designed by
John Ellis,
\begin{eqnarray}
  {\cal L}_{SM} &=& - \frac{1}{g^2} F_{\mu\nu}^2 + \bar{\psi} i {\not
    \!\! D} \psi + |D_\mu H|^2 - y \bar{\psi}\psi H +
  \frac{\theta}{64\pi^2} F \tilde{F} - \lambda (H^\dagger H)^2
  \nonumber \\
  & &+ \mu^2 H^\dagger H - \Lambda_{CC}.
\end{eqnarray}
The first line here has only dimensionless parameters and is only
logarithmically sensitive to the physics at $\Lambda_{UV}$.  On the
other hand, the last line has two parameters $\mu^2$ (mass dimension
2) and $\Lambda_{CC}$ (4) and {\it remember}\/ physics at
$\Lambda_{UV}$, the origin of the naturalness problems we will come
back to later.

On the other hand, the power-suppressed operators come in a great
variety.  For instance, those suppressed by two powers can be
\begin{equation}
  {\cal L}_6 = QQQL,\ \bar{L}\sigma^{\mu\nu}W_{\mu\nu}H l,\ 
  \epsilon_{abc}W_\nu^{a\mu}W_\lambda^{b\nu}W_\mu^{c\lambda},\
  (H^\dagger D_\mu H)^2,\ B_{\mu\nu} H^\dagger W^{\mu\nu} H, \cdots
\end{equation}
They may be seen in proton decay, $g_\mu -2$, triple gauge boson
vertex, $T$ and $S$-parameters in the precision EW observables,
respectively.  

It is interesting to note that there is actually only {\it one-type}\/
of operator we can write suppressed by a single power,
\begin{equation}
  {\cal L}_5 = (L H)(L H).
\end{equation}
After substituting the expectation values for the Higgs field, it is
nothing but the Majorana neutrino mass operator,
\begin{equation}
  \frac{1}{\Lambda_{UV}} {\cal L}_5 = \frac{v^2}{\Lambda_{UV}} \nu \nu .
\end{equation}
In other words, the neutrino mass can be viewed as the {\it leading order
  effect}\/ of the physics beyond the Standard Model!

The neutrino mass is actually a tiny effect.  Any kinematic effect is
suppressed by $m_\nu^2 / E_\nu^2 \approx (0.1{\rm eV}/{\rm GeV})^2
\approx 10^{-20}$!  Normally we don't think we can be sensitive to
such a small number in experiments.  However, there is one known
technique that is sensitive to very small numbers: {\it
  interferometry}\/, like in the Michaelson--Morley experiment.  For
this to be possible, there are three conditions: a coherent source, a
long baseline, and interference.  For some unknown reason, nature was
kind enough to provide us with all of the three necessary conditions.  Because
neutrinos interact so little, neutrinos maintain their coherence after
propagation over long distances, coming from the Sun, cosmic rays,
supernovae, accelerators, and reactors.  These sources are naturally
associated with rather long baselines.  And most remarkably,
significant interference effects require large mixing angles, which
happened to be the case with neutrinos!  Looking at it this way, it
may not be a huge surprise that the neutrino oscillation was the first
concrete evidence for physics beyond the Standard Model.  In other
words, the neutrino interferometry ({\it a.k.a.}\/ neutrino
oscillation) is a unique tool for studying physics at very high energies,
already probing up to $\Lambda \approx 10^{14}$~GeV.  Because of this
line of argument, I'm a big fan of neutrino physics--so much so that I
participated in the KamLAND experiment \cite{KamLAND} spending some
time building the detector, taking shifts, and serving on paper
committees.

One sometimes hears the criticism that flavor physics experiments,
quark, charged lepton, or neutrinos, have done little to improve our
understanding of the underlying physics, compared to those experiments
that focused on forces that led to the gauge theory and the
Englert--Brout--Higgs mechanism.  Indeed, we've known the pattern of
quark masses and mixing angles already for some time, with no clear
standard theory behind them.

We can, however, at least ask a question: does the patten of masses and
mixings require a new structure or symmetries beyond what we know?  I
claim that we can answer this question.

\begin{figure}
\begin{center}
\includegraphics[clip,width=.55\textwidth]{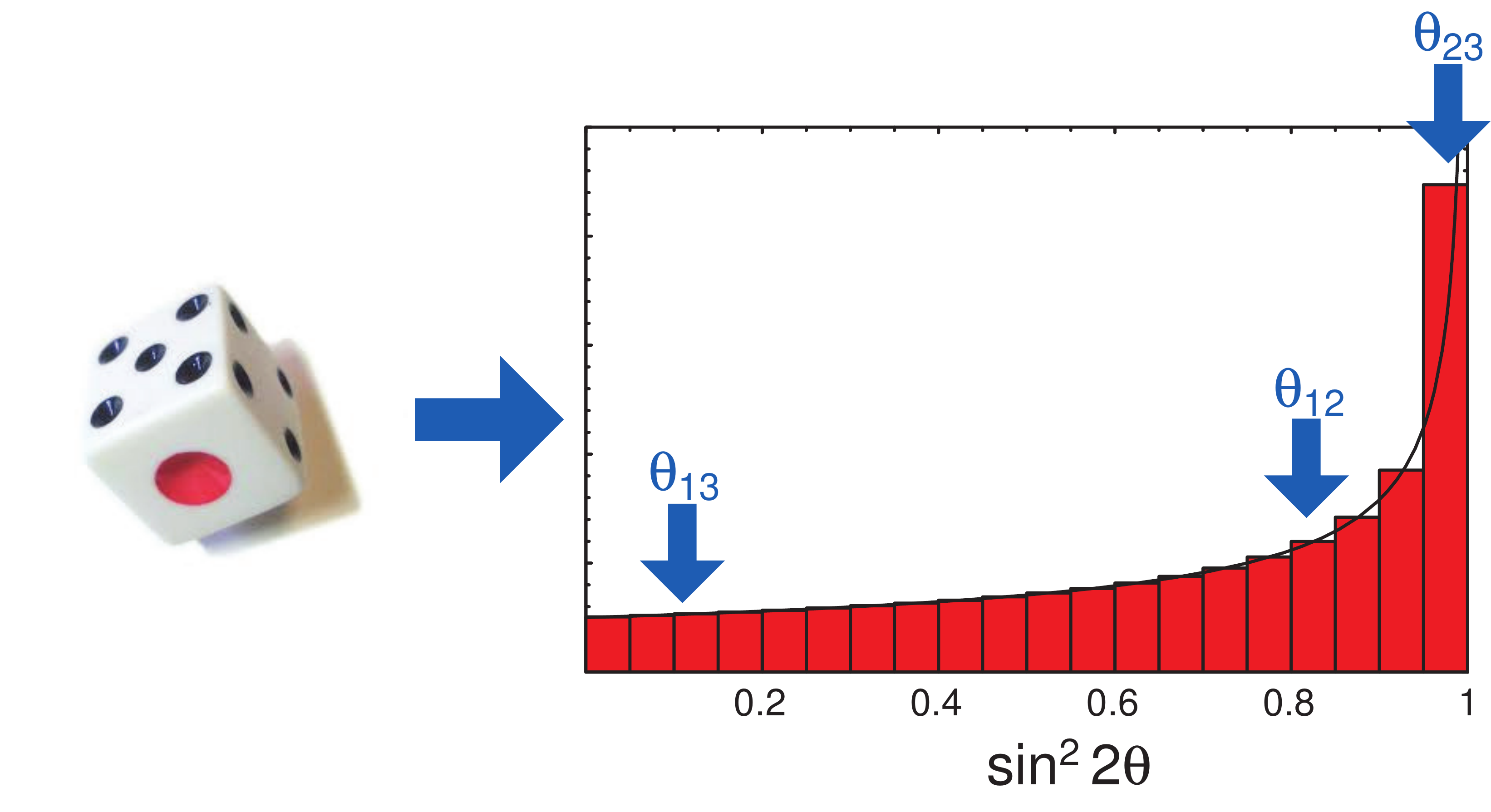}
\includegraphics[clip,width=.3\textwidth]{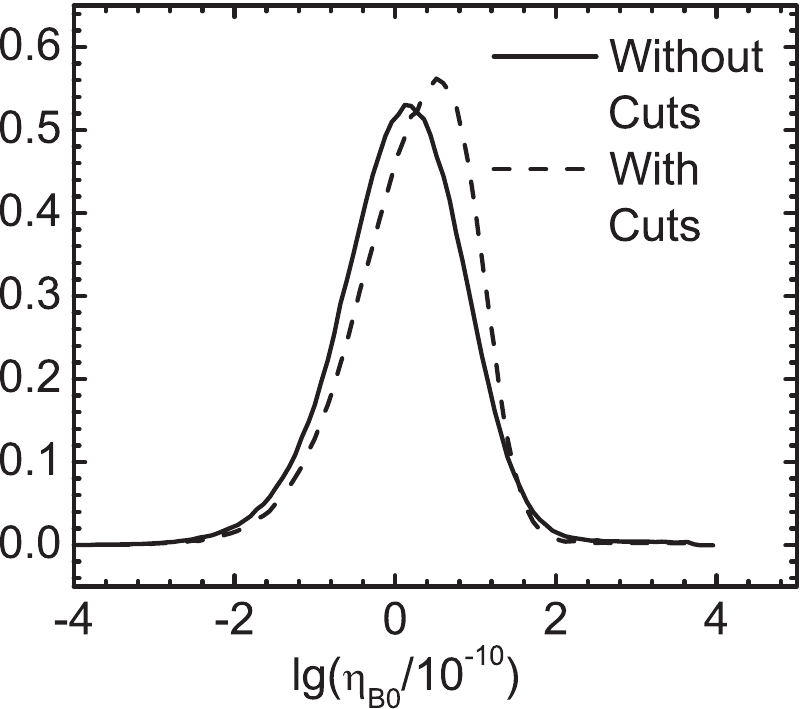}
\end{center}
\caption{Left: Prediction of random matrices for the neutrino mixing
  angles. Right: consequence of anarchy on the baryon asymmetry of the
  Universe via the thermal leptogenesis scenario.}
\label{fig:anarchy}
\end{figure}

I proposed the idea called {\it anarchy}\/, namely that the neutrino
masses and mixings do not require any new symmetries to be understood,
with Lawrence Hall and Neal Weiner \cite{Hall:1999sn}.  If there are
no symmetries or quantum numbers that distinguish three generations of
neutrinos, the neutrino mass matrix would consist of nearly random
comparable numbers without a particular structure.  We can understand
their consequences basically by throwing dice.  It actually leads to a
definite prediction for the mixing angles: the probability
distributions are given by the {\it Haar measure}\/ of the compact
groups, the unique measure that is invariant under both left- and
right-translations \cite{Haba:2000be}.  Then the distributions in
$x_{ij}=\sin^2 2\theta_{ij}$ turn out to be the same for all three
angles, $P(x) dx = \frac{1}{2} \frac{dx}{\sqrt{1-x}}$, peaked at the
maximal mixing $x=1$.  Given this, it is not surprising that the
neutrino oscillation was first discovered at the peak $x_{23} \approx
1$, then somewhere around the middle of the distribution $x_{12}
\approx 0.86$, and finally somewhat down the tail $x_{13} \approx
0.09$ (Fig.~\ref{fig:anarchy} Left).  In fact, Andr\'e de Gouv\^ea and
I did a Kolgomov--Smirnov test \cite{deGouvea:2003xe} to see if the
one draw made by nature is consistent with this probably distribution,
and found that it is 47\% probable \cite{deGouvea:2012ac}!  So we
learned indeed that the neutrino masses and mixings do not require any
deeper symmetries or new quantum numbers.  On the other hand, quarks
clearly do need additional input, which is yet to be understood.

The idea can be extended to the sector of right-handed neutrinos by
assuming that they have a hierarchy akin to those in the charged
leptons or quarks, $\epsilon^2 : \epsilon : 1$.  We take $\epsilon
\approx 0.1$.  With this structure, we can randomly generate the full
left- and right-handed neutrino mass matrices.  Xiaochuan Lu and I
identified that the Gaussian measure is the unique choice based on a
certain set of criteria, and found that the baryon asymmetry comes out
extremely well (Fig.~\ref{fig:anarchy} Right) \cite{Lu}.  This is
encouraging; in particular it is promising that the anarchy predicts
that the distribution in the CP-violating effect would peak at $\sin
\delta = \pm 1$ (or flat in $\delta$).

In fact, the CP violation in neutrino oscillation is the holy grail in
neutrino experiments currently being planned and discussed.  A possible CP
violation (assuming no matter effect) is given in terms of a product
of many factors,
\begin{eqnarray}
  P(\nu_\mu \rightarrow \nu_e) - P(\bar{\nu}_\mu \rightarrow \bar{\nu}_e)
  &=& -16 s_{12} c_{12} s_{13} c^2_{13} s_{23} c_{23} \nonumber \\
  & &
  \sin\delta \sin\frac{\Delta m_{12}^2 L}{4E}
  \sin\frac{\Delta m_{13}^2 L}{4E} \sin\frac{\Delta m_{23}^2 L}{4E}\ .
\end{eqnarray}
It is remarkable that all factors are now found to be large enough to
make this search feasible, the only unknown being the size of the CP
violation $\sin\delta$ itself.  Nature seems kind to us once again!
It was also interesting to learn at this symposium that beyond the
LBNE in the US and HyperK in Japan, there is a new discussion to use
the European Spallation Source in Sweden with a much shorter baseline
to look for CP violation \cite{Baussan:2013zcy}.

Coming back to dark matter, there is a big ongoing effort on Weakly
Interacting Massive Particle dark matter experiments from underground,
cosmic, and accelerator experiments \cite{Ong}.  If the bet is right,
we may see great discoveries sometime soon!

I argued that expeditions are needed to find where the next important
energy scale is to solve the five mysteries.  On the other hand,
so-called energy frontier experiments, namely those that rely on
high-energy colliders, target a rather specific energy scale.  This
leads us to ask the next question.

\section{Is the Energy Frontier Dead?}

The mantra in particle physics is to go to as high energy as technology
(and money) allows.  We are indeed going up a notch in 2015, restarting
the LHC at 13--14~TeV.  This is already exciting, improving reach on
new particles by a factor of two.  High-Luminosity LHC (HL-LHC) would
further improve the reach by 20--30\%.  We still have quite a bit of
room for discoveries.  More recently, there are discussions about a
potential 100~TeV $pp$ collider with a much bigger tunnel around CERN.

However, I see a problem arguing for the next much higher energy
machine now.  Given that the discovery of the Higgs boson made the
theory apparently complete, and the five mysteries I discussed have not
yet set particular energy scales, I don't know how we can justify the
energy of the next machine.  Does this mean that there is no case we can
make to build another high-energy collider?  Is Energy Frontier dead?

It remains true that the best argument we have right now to expect new
physics in the TeV range is the naturalness: we would
like to avoid fine-tuning between the bare $m_h^2$ and the radiative
correction (see, {\it e.g.}\/, \cite{Kolda:2000wi} for a plot).  Even
though many in the community are ditching the naturalness altogether,
I still take the argument seriously because it has worked many times
before.

One example I always bring up is the discovery of the positron
\cite{Murayama:1994kt,Murayama:2000dw}.  In classical electrodynamics,
the Coulomb self-energy of the electron is linearly divergent, $\Delta
m_e c^2 \sim \frac{e^2}{r_e}$, where $r_e$ is the ``size'' of the
electron.  It would have required a fine cancellation between the
``bare'' mass of the electron (which must be negative by the way) and
the correction to yield a small mass $m_e c^2 = 0.511$~MeV.  However,
the discovery of the positron and quantum mechanics told us that the
vacuum is always fluctuating, producing a pair of $e^+ e^-$, that
annihilates back to the vacuum within the time allowed by the
uncertainty principle $\Delta t \sim \hbar/\Delta E = \hbar/2m_e c^2$.
When you place an electron in this fluctuating vacuum, it may find a
(virtual) positron near it and decide to annihilate it.  Then the
other electron that was originally in the vacuum fluctuation is now
left out and becomes a ``real'' particle.  It turns out that this
process cancels the linear divergence exactly, leaving only a
logarithmic divergence $\Delta m_e c^2 = \frac{3\alpha}{4\pi} \log
\frac{\hbar}{m_e c r_e}$.  Even for an electron as small as the Planck
distance, it amounts to only 9\% correction.  The cancellation is
guaranteed by a (softly broken) chiral symmetry.  You can see that the
naturalness problem was solved by doubling the number of particles!

The idea of supersymmetry was pushed to repeat the history.  Because
the Higgs boson must repel itself, it also has a divergent
self-repulsion energy $\Delta m_H^2 \sim \lambda/r_H^2$.  But by
doubling the number of particles (namely introducing superpartners),
there is a cancellation between the self-repulsion among Higgs bosons,
and the induced attraction arising from the loop of higgsinos (fermionic
partner of the Higgs boson).  Again, the correction is down to a
logarithmic divergence, $\Delta m_H^2 \sim \frac{1}{(4\pi)^2}
m_{SUSY}^2 \log (\frac{\hbar}{m_H c r_H})$.

In the case of the electron, new physics (positron) appears ``early'' at
the Compton wave length $\hbar/m_e c \approx 400$~fm well before we
get down to the smaller ``classical radius of electron'' $r_c =
e^2/m_e c^2 \approx 1$~fm where the theory becomes fine-tuned.  In
another well-known case, however, nature did fine-tune it so that the
discovery was delayed.

The example is COBE (Cosmic Background Explorer) \cite{COBE} that
discovered the CMB anisotropy.  People expected anisotropy at the
level of $10^{-5}$ so that the observed large-scale structure can be
explained.  But the search went on, so much so that people started
writing articles questioning the inflationary cosmology itself.  When
COBE discovered the quadrupole moment, it was small.  Actually,
compared to our best prediction today based on the WMAP data, it was
nearly an order of magnitude smaller than theory.  This is usually
understood today as a consequence of {\it cosmic variance}\/, namely
that the quadrupole moment has only $2l+1=5$ numbers to be measured
and hence is subject to a statistical uncertainty of $O(1/\sqrt{5})$.
I find the observed quadrupole moment to be fine-tuned at the 2\%
level.

Note that the inflation was invented to solve the naturalness
problems, horizon problem and flatness problem of the standard Big
Bang cosmology.  It worked: the current data beautifully confirm
predictions of inflation.  But it was a little fine-tuned and it
required patience and more work.

So the moral I draw from these examples is that the naturalness
argument generally {\it does}\/ work.  But there are cases where
fine-tuning at the level of a few percent or even few per-mille (some
examples in nuclear physics are well-known, see \cite{Nima}).  Looking
back at Fig.~\ref{fig:finetuning}, we have not fully explored down to
that level of not-{\it that}\/-fine-tuning yet.  And it took ten years
for Tevatron to discover top.  Patience pays, hence my optimism.

But it is true that it is a slippery slope how much fine tuning we
tolerate.  Percent? Per-mille?  $10^{-6}$?  It is quite subjective or
matter of taste, and we cannot firmly set the energy for the next
collider based on this argument with any confidence.  Back to the
question again: is there any justification for the next high-energy
collider?

I believe there is.  The Higgs boson is the only newly discovered
particle at the LHC so far, but it is clearly an unusual particle.
And we know what energy is required to study it already, because we
know its mass.  We know where to look.

What is unusual about the particle discovered?  To the extent ATLAS
and CMS have managed to study it so far, it is consistent with
$J^{PC}=0^{++}$.  It has no spin!  We have never seen an elementary
spinless particle before, so this looks like a new breed.  Matter
particles like quarks and leptons come in three generations, and we
seem to understand their {\it context}\/.  That is, they look
familiar, they are siblings, and are a part of a big family.  Same
with the force particles.  They all belong to some kind of gauge
group.  Even though the groups differ, they follow the same principle.
Again, they have relatives, and have context.  But the Higgs boson is
totally different.  It is one of its kind, with no context within the
Standard Model.  That means that we introduced a kind of particle
nobody has seen before.  It is a faceless (spinless) intruder to our
happy family.  Yet it is supposed to do the most important job in the
theory.  The whole idea looks so artificial!

In fact, I couldn't believe the artificial aspect of the Standard
Model at all when I studied it back in grad school, so I became one of
the guilty ones to propose the Higgsless theories \cite{Csaki:2003dt}.
But this idea now appears {\it dead}\/.

Facing its existence, we are still puzzled by another question.  The
closest analogy we have in familiar systems is the superconductivity
we can study in the laboratory.  Instead of giving mass to $W$ and
$Z$-bosons, superconductivity gives mass to the photon.  In other
words, due to the Meissner effect, the magnetic field is repelled by
the superconductor, allowing it to enter only by a finite distance (the
penetration depth).  The magnetic field is short-ranged inside the
superconductor!  This case, however, is well understood.  It is caused
by the instability of Fermi surface when electrons are attracted to
each other by a weak force from the phonon exchange.  Cooper pairs
condense, making the magnetism short-ranged.  On the other hand, the
Standard Model does not tell us {\it why}\/ the Higgs boson condenses
in our Universe.  This is not only artificial, it is unsatisfying.

There are ideas to give context to the spinless Higgs boson.  There
may be many siblings and relatives.  The Higgs boson is just one among
the big spinless tribe, ome which happens to condense because of an
attractive force induced by the top-quark loops.  This idea is known
as supersymmetry.  An additional Higgs doublet is its sibling, and
there are many other spinless squarks and sleptons, that are its
relatives.  On the other hand, it may be composite, just like spinless
pions are made of spin 1/2 quarks.  In this case a new dynamics would
be required to bind the constituents together.  Or the Higgs boson may
actually be spinning, but if it does in extra dimensions we cannot
see, so we perceive it to be spinless.  In such a case, the Higgs
boson may actually be a gauge boson or even graviton.  These are all
familiar ideas we discussed for solving the naturalness problem.  Here
I'm not using the naturalness argument at all; but I still come back
to a similar set of ideas, namely that there are good reasons to
continue discussing these ideas.

Then, what should we do?  Of course, we should study this intruder as much as we
can!  If we look closely enough, maybe we can tell it it has siblings or
relatives.  We may find it has a finite size.  Or we may bring it back
to spin in our dimensions.  

Fortunately, the observed mass of 125~GeV is the best case scenario.
It allows us to measure branching fractions to $b\bar{b}$, $WW^*$,
$ZZ^*$, $gg$, $\tau^+\tau^-$, $c\bar{c}$, $\gamma\gamma$, $Z\gamma$,
possibly even $\mu^+ \mu^-$.  Some of them would not be accessible if
the Higgs were lighter or heavier by just a few tens of GeV.  It is
actually a dream case for experiments!

Looking back at the history of collider experiments, precision
measurements using leptons often revealed the next energy scale; we
went up there with hadrons, and we indeed found new things, which we
further studied with lepton probes.  One full cycle is the {\it
  precision measurements}\/ of neutral currents in polarized electron
deuteron scattering at SLAC.  The measured $\sin^2 \theta_W$ {\it
  predicted}\/ the masses of $W$ and $Z$.  Sp$\bar{\rm p}$S was built
to discover them, which indeed did.  After that LEP was built to study
them precisely and we {\it nailed}\/ the gauge sector of the Standard
Model.

The next cycle starts with LEP {\it predicting}\/ the top quark and
Higgs boson masses.  Tevatron and LHC were built for this purpose, and
as we know, thy did discover the predicted particles.  The obvious
thing to do next is to study them precisely to {\it nail}\/ the top
and Higgs sector at another lepton machine.

If the history is any guide, the future {\it precision measurement}\/
of the top and Higgs sector would tell us the next energy scale we
should go after.  We are on a scavenger hunt.  The Higgs boson
discovered is a {\it lamp post}\/, we need to look carefully at what's
under it, and hope to find a clue to the next destination.

Another reason why the precision study of the Higgs boson is exciting
is that the Higgs boson may be a {\it portal}\/ to a new sector
outside the Standard Model.  It may, for example, be a sector of the
dark matter particle.  To probe an operator ${\cal O}$ (with mass
dimension $d$) in the new sector, we need its coupling to the Standard
Model particles.  As we discussed before, all operators in the
Standard Model are of dimension four, except for the Higgs
mass-squared.  Therefore the couping is suppressed as {\it e.g.}\/,
$\frac{1}{\Lambda_{UV}^{d}} {\cal O} F_{\mu\nu}^2$, while the coupling
to the Higgs goes as $\frac{1}{\Lambda_{UV}^{d-2}} {\cal O} H^\dagger
H$.  Thus the coupling to the Higgs is enhanced by $\Lambda_{UV}^2$
relative to other operators.  The Higgs boson may be the window to the
new world.

In addition, once we build a new lepton collider to study Higgs and
top precisely, we can still hope that it discovers new particles
directly.  It is not true that LHC excluded everything below TeV.
Even a slepton of, say, 150~GeV is still allowed if it decays into a
neutralino heavier than 80~GeV or so.  LHC will improve limits to
heavier sleptons, but not much to close the gap when their masses are
close.

Given this, I'd think the strategy is clear.  We start with what we
have got.  We build a lepton collider that can study the top quark and
Higgs boson precisely.  This will be an evolutionary program, starting
with the $Zh$ threshold, measuring branching fractions and couplings
to $Z$, $W$, $b$, $c$, $\tau$, $g$, $\gamma$, even the decay into
invisible particles.  Then on to the $t\bar{t}$ threshold to study the
top quark compositeness, say, going up further in energies to make use
of new processes such as $WW$-fusion, $t\bar{t}h$ production for
$y_{tth}$, and multiple Higgs production for $\lambda_{hhh}$.  But we
should keep our eyes open to the possibility that we may also discover
new particles along the way.  Just in case we obtain a new piece of
information on new particles from the LHC, the lepton collider should
be extendable.  If we do see new particles, we should have the
capability of studying them in model-independent way, and to determine
their quantum numbers, spins, masses, and couplings.  The machine
should be one that we know how to build, so that we can propose it as soon
as an opportunity presents itself.

\begin{figure}
\begin{center}
\includegraphics[clip,width=.4\textwidth,angle=90]{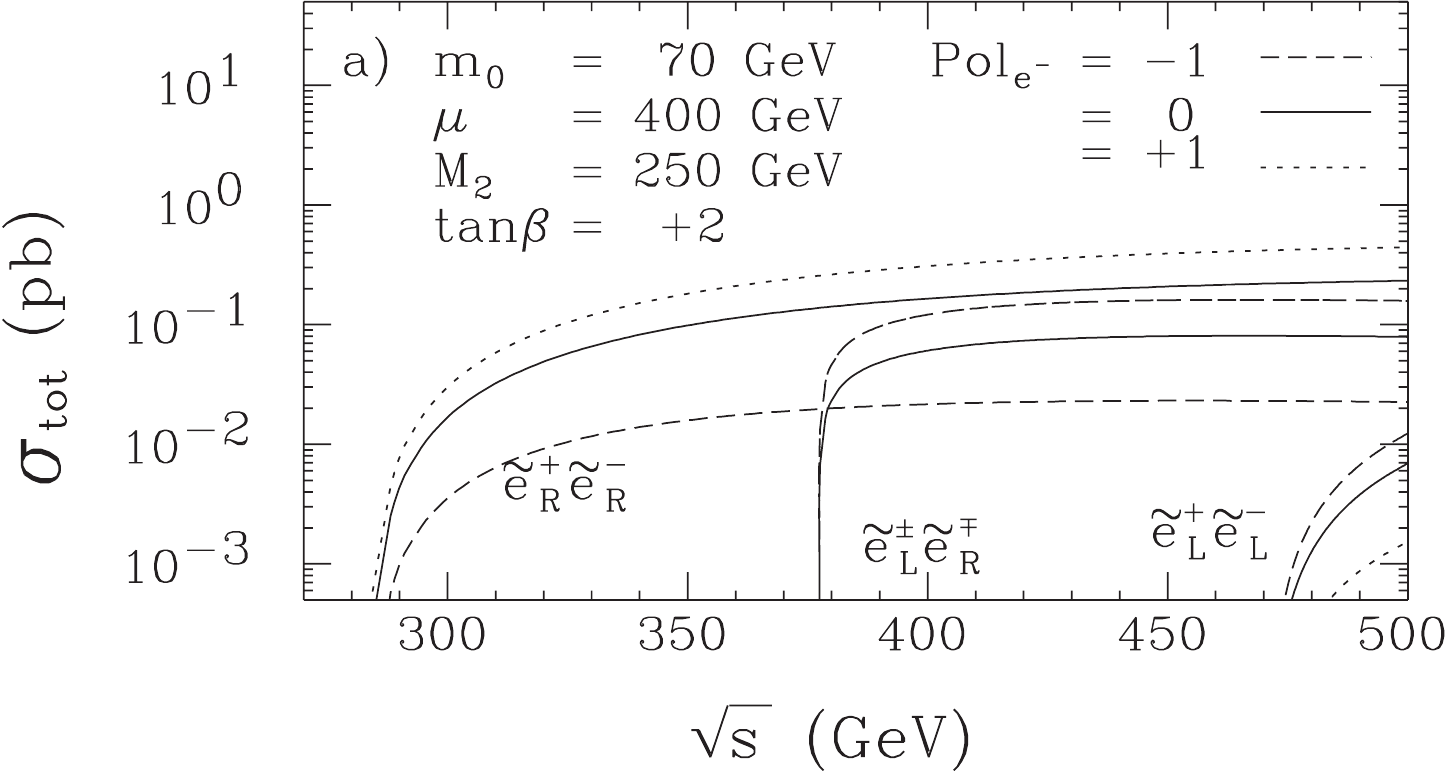}
\includegraphics[clip,width=.6\textwidth]{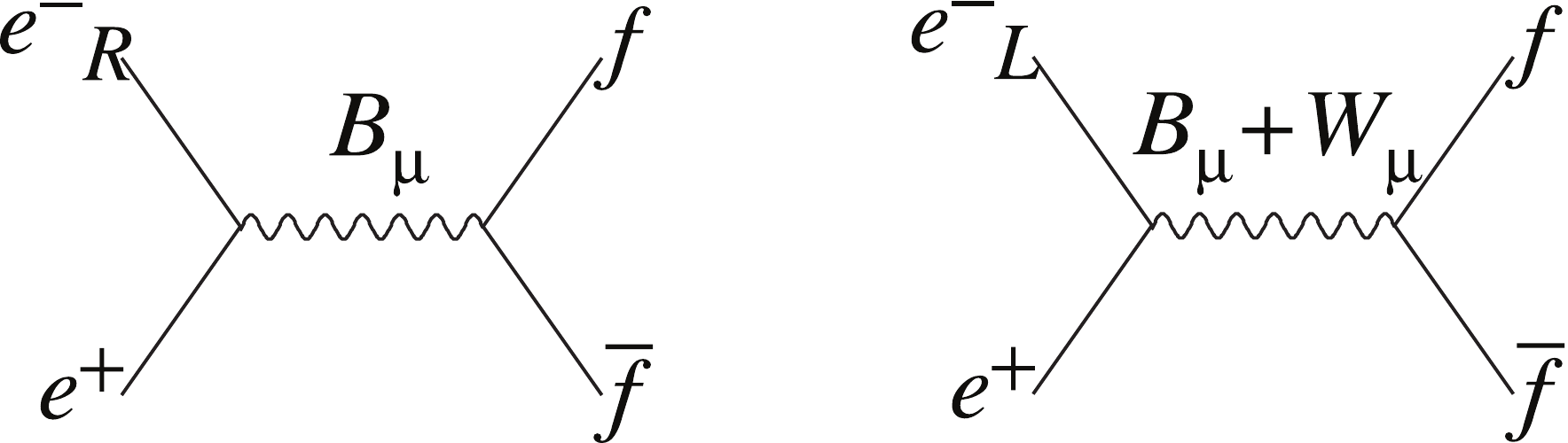}
\end{center}
\caption{Top: The cross sections for producing pairs of right-handed
  slectron vs left-handed selectron for various beam polarizations
  \cite{Tsukamoto:1993gt}.  Bottom: $s$-channcel production processes
  for new particles with polarized electrons.}
\label{fig:polarization}
\end{figure}

The planned International Linear Collider (ILC) fits this bill very
nicely, and its scientific case was judged very strong in the European
Strategy document adopted by CERN Council in May
\cite{EuropeanStrategy}.  The technology required for this is mature,
thanks to the Global Design Effort (GDE) led by Barry Barish that
finished the Technical Design Report (TDR) this year \cite{TDR}.  It
is extendable, so that we can increase the energy if needed and
affordable.  In addition, the longitudinal beam polarization provides
a crucial tool.  The bottom diagrams in Fig.~\ref{fig:polarization}
show that different electron polarization has different gauge bosons
in the $s$-channel.  At these energies, we can neglect $m_Z^2 \ll s$
as an approximation.  Then the gauge bosons exchanged are either
$U(1)_Y$ gauge boson $B_\mu$, or the neutral $SU(2)_L$ gauge boson
$W_\mu^0$.  The right-handed electron is a singlet under $SU(2)_L$
(the subscript $L$ stands for left-handed), and does not couple to $W_\mu^0$.
Therefore, the $s$-channel production goes as $|g' Y_f|^2$ and
directly measures the hypercharge $Y_f$ of the new particle $f$.  On
the other hand, the left-handed electron couples to both, and the
cross section goes as $|g' Y_f + g I_{3f}|^2$.  Knowing $Y_f$, we can
determine $I_{3f}$ model-independently.  One can see how much the
cross sections vary depending on the beam polarization in the top plot
of Fig.~\ref{fig:polarization}.

There are many studies of spin and mass measurements, which can be
done even when particles decay into invisible particles.  Combination
of the LHC and ILC data may allow us to even compute the cosmological
abundance of the invisible particle, possibly verifying that it is the
dark matter of the Universe \cite{Baltz:2006fm}.

But isn't ILC too expensive to be ever built?  Through some miracle,
many politicians in Japan are interested in hosting the ILC as a
global project.  They would like to open up the country to talented,
intellectual people from abroad.  They would like to find prestige in
hosting highly visible large international project.  They also want to
use the ILC to build up infrastructure, a technological base, and they
hope to find economic benefits.  More than 20\% of the Diet members
signed up to support the ILC, in a group named {\it Federation of Diet
  Members for Promotion of the ILC}\/.  When Lyn Evans visited Japan
in March, the prime minister Shinzo Abe agreed to meet him \cite{Lyn},
and he said that he appreciated the significance of the ILC as ``a
dream for humankind.''  His opening address in the 183rd session of
the Diet mentioned advanced accelerator technology as one of the
innovation areas in which he wants Japan to excel \cite{address}.
There are many industry associations actively supporting the ILC; the
media is highly interested as well.  And the discovery of the Higgs
boson has fueled interest even further.  I'd think there is a high
enough level of interest for the Japanese government to initiate
discussions with other potentially interested countries to form an
international framework for a global ILC project hosted in Japan.  I
am not absolutely sure; but it doesn't look impossible so far.

Having discussed expeditions to high-energy scales, and precision
studis of the Higgs and top to identify the next energy scale(s),
there are plenty of things that we can and will do in the near future
in our field.  However, it still leaves one question that has been
haunting me.

\section{Do we ever understand Dark Energy?}

Dark Energy is such a big mystery that I cannot gauge how we may ever
understand it.  Does this mean that it is useless to try to measure
its properties precisely?

I don't know.  But all I can say is that a percent-level or better
measurement is what I consider {\it precise}\/.  If the equation of
state parameter $w\equiv p/\rho =0 \pm 0.01$, I may give up and say it
is the cosmological constant, accidentally small in a landscape of
$10^{500}$ universes.  But it may turn out to be $w=0.05 \pm 0.01$,
pointing us in a new direction.  I believe that it is worth the try.

\begin{figure}
\begin{center}
\includegraphics[clip,width=.9\textwidth]{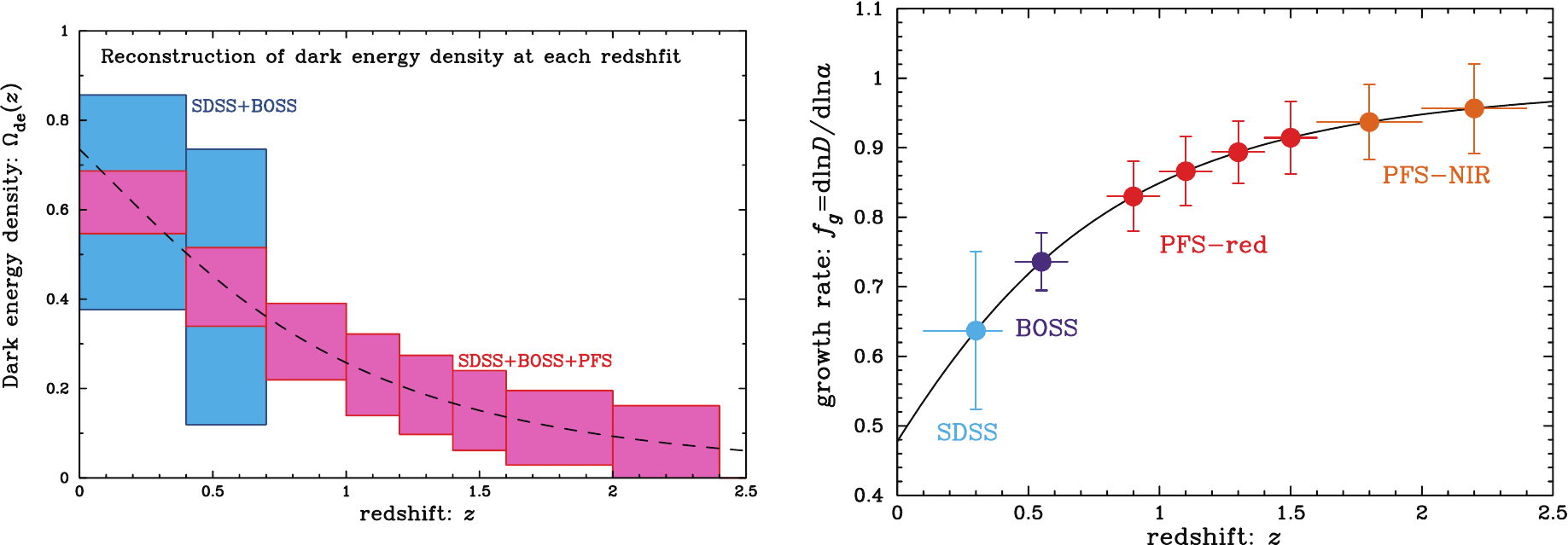}
\end{center}
\caption{Left: Projected accuracy in measuring $\Omega_{DE} (z)$ using
  PFS on Subaru telescope.  Right: Test of general relativity on
  cosmological distances using the growth rate of structure.  Both
  taken from \cite{Ellis:2012rn}.}
\label{fig:PFS}
\end{figure}

I lead a major dark energy experiment called SuMIRe (Subaru
Measurement of Images and Redshifts) \cite{SuMIRe}.  It combines
imaging and spectroscopy on the 8.2m-diameter Subaru telescope, a
major step up from the wildly successful SDSS (Sloan Digital Sky
Survey) \cite{SDSS}.  The first stage is the approved imaging survey
with Hyper Suprime-Cam for 300 nights, with nearly 0.9 billion pixels
with a field of view of 1.7 square degrees.  It will image hundreds of
millions of galaxies.  The next stage is a spectroscopic survey with
the Prime Focus Spectrograph for (hopefully) 300 nights, with 2400
optical fibers, controlled robotically, being targeted at galaxies
chosen from the imaging survey.  For instance, it will yield a
model-independent measurement of the evolution of the dark energy
fractions as a function of the redshift (left Fig.~\ref{fig:PFS}) and
provide a test of general relativity at cosmological distances
(right Fig.~\ref{fig:PFS}).

We should do what we can do, and we will see what we find!

\appendix

\section{Pushing back on fine-tuning \label{sec:appendix}}

Take supersymmetry.  There are two issues facing the experimental
data.  The first one is that the mass of the discovered Higgs boson
$m_h$ is a little too high within the MSSM, which predicts $m_h \leq
m_Z$ at the tree-level.  Even though the Higgs mass can be pushed up
by the radiative correction as
\begin{equation}
  m_h^2 \simeq m_Z^2 
  + \frac{3}{4\pi} h_t^4 v^2 \log
  \frac{m_{\tilde{t}_1}m_{\tilde{t}_2}}{m_t^2},
\end{equation}
it would require a large scalar top mass, which would feed into the
radiative correction to the Higgs mass-squared term
\begin{equation}
  \Delta m_{H_u}^2 \simeq - 12 \frac{h_t^2}{16\pi^2} m_{\tilde{t}}^2
  \log \frac{\Lambda_{UV}}{\mu_{IR}}\ .
\end{equation}
Therefore, a larger physical Higgs mass in the MSSM indirectly implies
exponentially worse fine-tuning in $m_{H_u}^2$ between the bare
parameter in the Lagrangian and the radiative correction above.

\begin{figure}
\begin{center}
\includegraphics[clip,width=.5\textwidth]{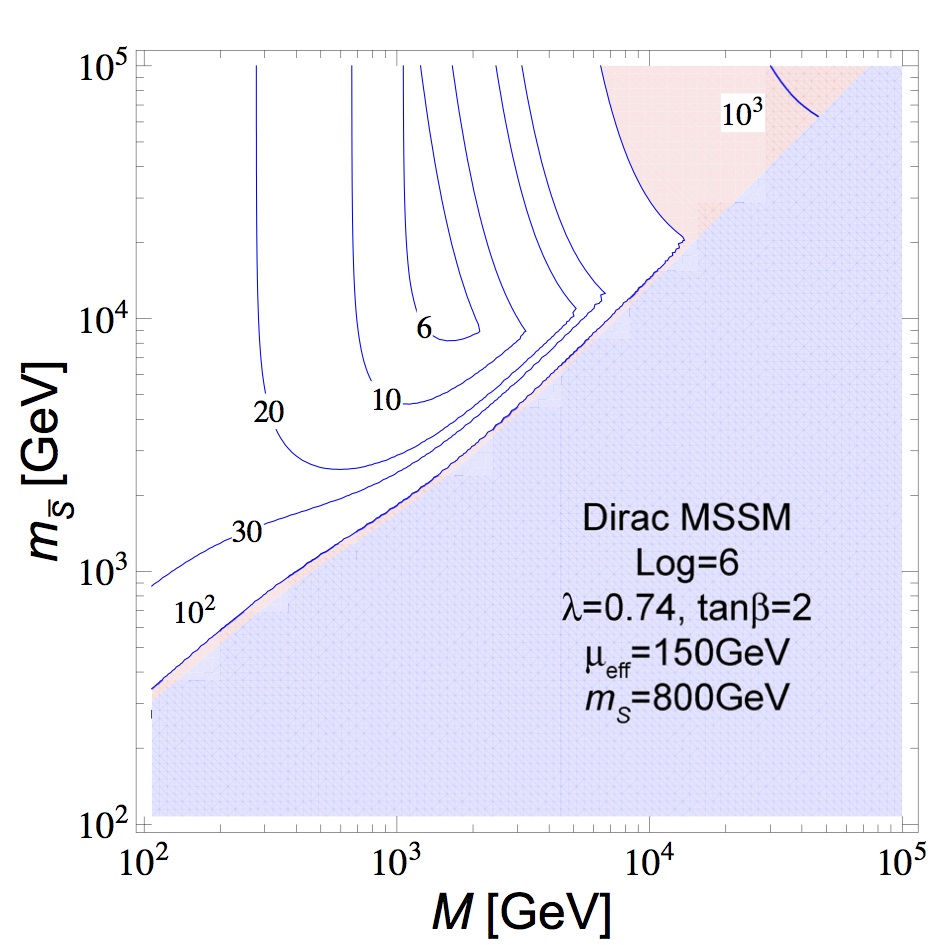}
\end{center}
\caption{Degree of fine-tuning in Dirac NMSSM that shows it can be as
  good as a factor of six even when $m_{\bar{S}} \gg m_h$.  Taken from
  \cite{Lu:2013cta}.}
\label{fig:DiracNMSSM}
\end{figure}

This can be avoided if there is an additional contribution to the
Higgs self-coupling, such as in the massive NMSSM $W = \lambda S H_u
H_d + \frac{1}{2} M S^2$.  However the contribution decouples in the
limit $M \gg m_h$ as
\begin{equation}
  \lambda^2 = 1 - \frac{M^2}{M^2 + m_S^2}\ .
\end{equation}
This can be prevented for a large soft mass $m_S^2 \gg M^2$ as a
non-decoupling effect \cite{Dine:2007xi}, but $m_S^2$ then feeds into
$\Delta m_{H_u}^2 = \frac{1}{16\pi^2} 2 \lambda^2 m_S^2 \log
\frac{\Lambda_{UV}}{\mu_{IR}}$, re-introducing the fine-tuning.

Together with Xiaochuan Lu, Josh Ruderman, and Kohsaku Tobioka, we've
come up with an idea that we call {\it semi-soft supersymmetry
  breaking}\/ \cite{Lu:2013cta}.  Using the NMSSM (Next-to-Minimal
Supersymmetric Standard Model) of the Dirac-type, $W=\lambda S H_u H_d
+ M S \bar{S}$, the singlet field $\bar{S}$ couples to the rest of the
model only through a dimensionful coupling $M$.  It can then be proven
that the limit $m_{\bar{S}}^2 \rightarrow \infty$ does not
re-introduce fine-tuning (Fig.~\ref{fig:DiracNMSSM}) even though it
looks like a hard breaking of supersymmetry, hence {\it semi-soft}\/.

The second problem with supersymmetry is its non-observation in direct
searches.  It is well-known that a quasi-degenerate spectrum among
supersymmetric particles makes the search difficult because of small
$Q$-values in decays and hence small ${\not\!\! E}_T$ (see, {\it
  e.g.}\/, \cite{LeCompte:2011fh}).  However, such a spectrum lacked
theoretical motivation: in particular, why should scalars and gauginos
be degenerate?

Together with Yasunori Nomura, Satoshi Shirai, Kohsaku Tobioka, I
proposed that supersymmetry broken by boundary conditions in extra
dimensions would automatically give the same mass to {\it all}\/
gauginos and sfermions at the tree-level split only by loop effects
\cite{Murayama:2012jh}, similar to the Universal Extra Dimension (UED)
\cite{Cheng:2002ab}.  Correspondingly, the experimental limit is
weaker.  A dedicated search with ISR should improve the limit like in
the UED case \cite{Murayama:2011hj}.

This work was supported in part by the U.S. DOE under Contract
No.~DEAC03-76SF00098, by the NSF under Grant No.~PHY-1002399, by the
JSPS Grant (C) No.~23540289, by the FIRST program Subaru Measurements
of Images and Redshifts (SuMIRe), CSTP, and by WPI, MEXT, Japan.


\section*{References}
\begin{thebibliography}{10}

%\cite{Aad:2012tfa}
\bibitem{Aad:2012tfa} 
  G.~Aad {\it et al.}  [ATLAS Collaboration],
  %``Observation of a new particle in the search for the Standard Model Higgs boson with the ATLAS detector at the LHC,''
  Phys.\ Lett.\ B {\bf 716}, 1 (2012).
%   [arXiv:1207.7214 [hep-ex]].
  %%CITATION = ARXIV:1207.7214;%%
  %1654 citations counted in INSPIRE as of 01 Oct 2013

%\cite{Chatrchyan:2012ufa}
\bibitem{Chatrchyan:2012ufa} 
  S.~Chatrchyan {\it et al.}  [CMS Collaboration],
  %``Observation of a new boson at a mass of 125 GeV with the CMS experiment at the LHC,''
  Phys.\ Lett.\ B {\bf 716}, 30 (2012).
%   [arXiv:1207.7235 [hep-ex]].
  %%CITATION = ARXIV:1207.7235;%%
  %1624 citations counted in INSPIRE as of 01 Oct 2013

%\cite{Ade:2013uln}
\bibitem{Ade:2013uln} 
  P.~A.~R.~Ade {\it et al.}  [Planck Collaboration],
  %``Planck 2013 results. XXII. Constraints on inflation,''
  arXiv:1303.5082 [astro-ph.CO].
  %%CITATION = ARXIV:1303.5082;%%
  %219 citations counted in INSPIRE as of 01 Oct 2013

%\cite{Afonso:2002xq}
\bibitem{Afonso:2002xq} 
  C.~Afonso {\it et al.}  [EROS Collaboration],
  %``Limits on galactic dark matter with 5 years of EROS SMC data,''
  Astron.\ Astrophys.\  {\bf 400}, 951 (2003).
%   [astro-ph/0212176].
  %%CITATION = ASTRO-PH/0212176;%%
  %85 citations counted in INSPIRE as of 01 Oct 2013

%\cite{Spergel:2003cb}
\bibitem{Spergel:2003cb} 
  D.~N.~Spergel {\it et al.}  [WMAP Collaboration],
  %``First year Wilkinson Microwave Anisotropy Probe (WMAP) observations: Determination of cosmological parameters,''
  Astrophys.\ J.\ Suppl.\  {\bf 148}, 175 (2003).
%   [astro-ph/0302209].
  %%CITATION = ASTRO-PH/0302209;%%
  %6995 citations counted in INSPIRE as of 01 Oct 2013

%\cite{Hu:2000ke}
\bibitem{Hu:2000ke} 
  W.~Hu, R.~Barkana and A.~Gruzinov,
  %``Cold and fuzzy dark matter,''
  Phys.\ Rev.\ Lett.\  {\bf 85}, 1158 (2000).
%   [astro-ph/0003365].
  %%CITATION = ASTRO-PH/0003365;%%
  %182 citations counted in INSPIRE as of 01 Oct 2013

%\cite{Fukuda:1998mi}
\bibitem{Fukuda:1998mi} 
  Y.~Fukuda {\it et al.}  [Super-Kamiokande Collaboration],
  %``Evidence for oscillation of atmospheric neutrinos,''
  Phys.\ Rev.\ Lett.\  {\bf 81}, 1562 (1998).
%   [hep-ex/9807003].
  %%CITATION = HEP-EX/9807003;%%
  %4006 citations counted in INSPIRE as of 01 Oct 2013

%\cite{Ahmad:2002jz}
\bibitem{Ahmad:2002jz} 
  Q.~R.~Ahmad {\it et al.}  [SNO Collaboration],
  %``Direct evidence for neutrino flavor transformation from neutral current interactions in the Sudbury Neutrino Observatory,''
  Phys.\ Rev.\ Lett.\  {\bf 89}, 011301 (2002).
%   [nucl-ex/0204008].
  %%CITATION = NUCL-EX/0204008;%%
  %2214 citations counted in INSPIRE as of 01 Oct 2013

%\cite{Eguchi:2002dm}
\bibitem{Eguchi:2002dm} 
  K.~Eguchi {\it et al.}  [KamLAND Collaboration],
  %``First results from KamLAND: Evidence for reactor anti-neutrino disappearance,''
  Phys.\ Rev.\ Lett.\  {\bf 90}, 021802 (2003).
%   [hep-ex/0212021].
  %%CITATION = HEP-EX/0212021;%%
  %2034 citations counted in INSPIRE as of 01 Oct 2013

%\cite{An:2012eh}
\bibitem{An:2012eh} 
  F.~P.~An {\it et al.}  [DAYA-BAY Collaboration],
  %``Observation of electron-antineutrino disappearance at Daya Bay,''
  Phys.\ Rev.\ Lett.\  {\bf 108}, 171803 (2012).
%   [arXiv:1203.1669 [hep-ex]].
  %%CITATION = ARXIV:1203.1669;%%
  %671 citations counted in INSPIRE as of 01 Oct 2013

%\cite{Perlmutter:1998np}
\bibitem{Perlmutter:1998np} 
  S.~Perlmutter {\it et al.}  [Supernova Cosmology Project Collaboration],
  %``Measurements of Omega and Lambda from 42 high redshift supernovae,''
  Astrophys.\ J.\  {\bf 517}, 565 (1999).
%   [astro-ph/9812133].
  %%CITATION = ASTRO-PH/9812133;%%
  %7216 citations counted in INSPIRE as of 01 Oct 2013

%\cite{Riess:1998cb}
\bibitem{Riess:1998cb} 
  A.~G.~Riess {\it et al.}  [Supernova Search Team Collaboration],
  %``Observational evidence from supernovae for an accelerating universe and a cosmological constant,''
  Astron.\ J.\  {\bf 116}, 1009 (1998).
%   [astro-ph/9805201].
  %%CITATION = ASTRO-PH/9805201;%%
  %6816 citations counted in INSPIRE as of 01 Oct 2013

%\cite{Abe:2001xe}
\bibitem{Abe:2001xe} 
  K.~Abe {\it et al.}  [Belle Collaboration],
  %``Observation of large CP violation in the neutral $B$ meson system,''
  Phys.\ Rev.\ Lett.\  {\bf 87}, 091802 (2001).
%   [hep-ex/0107061].
  %%CITATION = HEP-EX/0107061;%%
  %669 citations counted in INSPIRE as of 01 Oct 2013

%\cite{Aubert:2001nu}
\bibitem{Aubert:2001nu} 
  B.~Aubert {\it et al.}  [BaBar Collaboration],
  %``Observation of CP violation in the $B^0$ meson system,''
  Phys.\ Rev.\ Lett.\  {\bf 87}, 091801 (2001).
%   [hep-ex/0107013].
  %%CITATION = HEP-EX/0107013;%%
  %644 citations counted in INSPIRE as of 01 Oct 2013

%\cite{Jarlskog:1985ht}
\bibitem{Jarlskog:1985ht} 
  C.~Jarlskog,
  %``Commutator of the Quark Mass Matrices in the Standard Electroweak Model and a Measure of Maximal CP Violation,''
  Phys.\ Rev.\ Lett.\  {\bf 55}, 1039 (1985).
  %%CITATION = PRLTA,55,1039;%%
  %1102 citations counted in INSPIRE as of 01 Oct 2013

\bibitem{Ligeti} I thank Zoltan Ligeti for the concept for this
  schematic plot.

\bibitem{KamLAND} Kamioka Liquid scintillator Anti-Neutrino Detector
  \url{http://www.awa.tohoku.ac.jp/kamlande/}

%\cite{Hall:1999sn}
\bibitem{Hall:1999sn} 
  L.~J.~Hall, H.~Murayama and N.~Weiner,
  %``Neutrino mass anarchy,''
  Phys.\ Rev.\ Lett.\  {\bf 84}, 2572 (2000).
%   [hep-ph/9911341].
  %%CITATION = HEP-PH/9911341;%%
  %195 citations counted in INSPIRE as of 01 Oct 2013

%\cite{Haba:2000be}
\bibitem{Haba:2000be} 
  N.~Haba and H.~Murayama,
  %``Anarchy and hierarchy,''
  Phys.\ Rev.\ D {\bf 63}, 053010 (2001).
%   [hep-ph/0009174].
  %%CITATION = HEP-PH/0009174;%%
  %137 citations counted in INSPIRE as of 01 Oct 2013

%\cite{deGouvea:2003xe}
\bibitem{deGouvea:2003xe} 
  A.~de Gouv\^ea and H.~Murayama,
  %``Statistical test of anarchy,''
  Phys.\ Lett.\ B {\bf 573}, 94 (2003).
%   [hep-ph/0301050].
  %%CITATION = HEP-PH/0301050;%%
  %74 citations counted in INSPIRE as of 01 Oct 2013

%\cite{deGouvea:2012ac}
\bibitem{deGouvea:2012ac} 
  A.~de Gouv\^ea and H.~Murayama,
  %``Neutrino Mixing Anarchy: Alive and Kicking,''
  arXiv:1204.1249 [hep-ph].
  %%CITATION = ARXIV:1204.1249;%%
  %27 citations counted in INSPIRE as of 01 Oct 2013

\bibitem{Lu}
  Xiaochuan Lu and Hitoshi Murayama, in preparation.

%\cite{Baussan:2013zcy}
\bibitem{Baussan:2013zcy} 
  E.~Baussan {\it et al.}  [ESSnuSB Collaboration],
  %``A Very Intense Neutrino Super Beam Experiment for Leptonic CP Violation Discovery based on the European Spallation Source Linac: A Snowmass 2013 White Paper,''
  arXiv:1309.7022 [hep-ex].
  %%CITATION = ARXIV:1309.7022;%%
  %3 citations counted in INSPIRE as of 01 Dec 2013

\bibitem{Ong}
Ren\'e Ong, in this proceedings.

%\cite{Kolda:2000wi}
\bibitem{Kolda:2000wi} 
  C.~F.~Kolda and H.~Murayama,
  %``The Higgs mass and new physics scales in the minimal standard model,''
  JHEP {\bf 0007}, 035 (2000).
%   [hep-ph/0003170].
  %%CITATION = HEP-PH/0003170;%%
  %87 citations counted in INSPIRE as of 01 Oct 2013

%\cite{Murayama:1994kt}
\bibitem{Murayama:1994kt} 
  H.~Murayama,
  %``Supersymmetry,''
  hep-ph/9410285.
  %%CITATION = HEP-PH/9410285;%%
  %13 citations counted in INSPIRE as of 01 Oct 2013

%\cite{Murayama:2000dw}
\bibitem{Murayama:2000dw} 
  H.~Murayama,
  %``Supersymmetry phenomenology,''
  hep-ph/0002232.
  %%CITATION = HEP-PH/0002232;%%
  %34 citations counted in INSPIRE as of 01 Oct 2013

\bibitem{COBE} Cosmic Background Explorer \url{http://lambda.gsfc.nasa.gov/product/cobe/}

\bibitem{Nima} Nima Arkani-Hamed in this proceedings.

%\cite{Csaki:2003dt}
\bibitem{Csaki:2003dt} 
  C.~Csaki, C.~Grojean, H.~Murayama, L.~Pilo and J.~Terning,
  %``Gauge theories on an interval: Unitarity without a Higgs,''
  Phys.\ Rev.\ D {\bf 69}, 055006 (2004)
  [hep-ph/0305237].
  %%CITATION = HEP-PH/0305237;%%
  %436 citations counted in INSPIRE as of 31 Dec 2013

\bibitem{EuropeanStrategy}
\url{http://council.web.cern.ch/council/en/EuropeanStrategy/ESParticlePhysics.html}

\bibitem{TDR}
  \url{http://www.linearcollider.org/ILC/Publications/Technical-Design-Report}

%\cite{Baltz:2006fm}
\bibitem{Baltz:2006fm} 
  E.~A.~Baltz, M.~Battaglia, M.~E.~Peskin and T.~Wizansky,
  %``Determination of dark matter properties at high-energy colliders,''
  Phys.\ Rev.\ D {\bf 74}, 103521 (2006).
%   [hep-ph/0602187].
  %%CITATION = HEP-PH/0602187;%%
  %219 citations counted in INSPIRE as of 01 Oct 2013

%\cite{Tsukamoto:1993gt}
\bibitem{Tsukamoto:1993gt} 
  T.~Tsukamoto, K.~Fujii, H.~Murayama, M.~Yamaguchi and Y.~Okada,
  %``Precision study of supersymmetry at future linear e+ e- colliders,''
  Phys.\ Rev.\ D {\bf 51}, 3153 (1995).
  %%CITATION = PHRVA,D51,3153;%%
  %177 citations counted in INSPIRE as of 31 Dec 2013

\bibitem{Lyn}
  \url{http://www.kantei.go.jp/jp/96_abe/actions/201303/27ilc_hyokei.html}

\bibitem{address} \url{http://www.kantei.go.jp/jp/96_abe/statement2/20130228siseuhousin.html}

\bibitem{SuMIRe} Subaru Measurements of Images and Redshifts 
\url{http://sumire.ipmu.jp/en/}

\bibitem{SDSS} Sloan Digital Sky Survey \url{http://www.sdss.org}

%\cite{Ellis:2012rn}
\bibitem{Ellis:2012rn} 
  R.~Ellis {\it et al.}  [PFS Team Collaboration],
  %``Extragalactic Science and Cosmology with the Subaru Prime Focus Spectrograph (PFS),''
  arXiv:1206.0737 [astro-ph.CO].
  %%CITATION = ARXIV:1206.0737;%%
  %31 citations counted in INSPIRE as of 01 Oct 2013

%\cite{Dine:2007xi}
\bibitem{Dine:2007xi} 
  M.~Dine, N.~Seiberg and S.~Thomas,
  %``Higgs physics as a window beyond the MSSM (BMSSM),''
  Phys.\ Rev.\ D {\bf 76}, 095004 (2007).
%   [arXiv:0707.0005 [hep-ph]].
  %%CITATION = ARXIV:0707.0005;%%

%\cite{Lu:2013cta}
\bibitem{Lu:2013cta} 
  X.~Lu, H.~Murayama, J.~T.~Ruderman and K.~Tobioka,
  %``A Natural Higgs Mass in Supersymmetry from Non-Decoupling Effects,''
  arXiv:1308.0792 [hep-ph].
  %%CITATION = ARXIV:1308.0792;%%
  %2 citations counted in INSPIRE as of 01 Oct 2013

%\cite{LeCompte:2011fh}
\bibitem{LeCompte:2011fh} 
  T.~J.~LeCompte and S.~P.~Martin,
  %``Compressed supersymmetry after 1/fb at the Large Hadron Collider,''
  Phys.\ Rev.\ D {\bf 85}, 035023 (2012).
%   [arXiv:1111.6897 [hep-ph]].
  %%CITATION = ARXIV:1111.6897;%%
  %45 citations counted in INSPIRE as of 01 Oct 2013

%\cite{Murayama:2012jh}
\bibitem{Murayama:2012jh} 
  H.~Murayama, Y.~Nomura, S.~Shirai and K.~Tobioka,
  %``Compact Supersymmetry,''
  Phys.\ Rev.\ D {\bf 86}, 115014 (2012).
%   [arXiv:1206.4993 [hep-ph]].
  %%CITATION = ARXIV:1206.4993;%%
  %12 citations counted in INSPIRE as of 01 Oct 2013

%\cite{Cheng:2002ab}
\bibitem{Cheng:2002ab} 
  H.~-C.~Cheng, K.~T.~Matchev and M.~Schmaltz,
  %``Bosonic supersymmetry? Getting fooled at the CERN LHC,''
  Phys.\ Rev.\ D {\bf 66}, 056006 (2002).
%   [hep-ph/0205314].
  %%CITATION = HEP-PH/0205314;%%
  %305 citations counted in INSPIRE as of 01 Oct 2013

%\cite{Murayama:2011hj}
\bibitem{Murayama:2011hj} 
  H.~Murayama, M.~M.~Nojiri and K.~Tobioka,
  %``Improved discovery of a nearly degenerate model: MUED using MT2 at the LHC,''
  Phys.\ Rev.\ D {\bf 84}, 094015 (2011).
%   [arXiv:1107.3369 [hep-ph]].
  %%CITATION = ARXIV:1107.3369;%%
  %18 citations counted in INSPIRE as of 01 Oct 2013

\end{thebibliography}
\end{document}